\documentstyle[prc,preprint,tighten,aps]{revtex}
\begin{document}
\draft
\title{Low-energy ${\bf M1}$ strength in the 
$\bbox{^7}$Li($\bbox{p,\gamma_0}$)$\bbox{^8}$Be 
reaction} 
\author{Attila Cs\'ot\'o$^{1,2}$ and Steven Karataglidis$^1$}
\address{$^1$National Superconducting Cyclotron Laboratory,
Michigan State University, East Lansing, Michigan 48824 \\
$^2$Theoretical Division, Los Alamos National Laboratory,
Los Alamos, New Mexico 87545}
\date{\today}

\maketitle

\begin{abstract}
The $^7$Li($p,\gamma_0$)$^8$Be reaction is studied in a microscopic
cluster model. All relevant subsystem properties are well
reproduced. The calculated astrophysical $S$ factor is in good
agreement with the experimental data, although some $M1$ strength is
missing in the $1^+;1$ resonance region. There is no contribution from
the $1^+;0$ state. We estimate an $M1$ strength of 3.5\% at 80~keV. By
assuming the experimental resonance parameters and channel spin ratio
for the first $1^+$ state that $M1$ strength is 6.3\%, in good
agreement with an $R$ matrix analysis. The large (20\%) $M1$ strength
determined from a transition matrix element analysis may be caused by
assuming a vanishing $0p_{3/2}$ spectroscopic factor, which is not
supported by shell model calculations.
\end{abstract}

\pacs{{\em Keywords}: NUCLEAR REACTIONS $^7$Li($p,\gamma_0$)$^8$Be, 
solar neutrinos, cluster model, shell model \\
{\em PACS}: 25.40.Lw; 26.65.+t; 21.60.Gx; 21.45.+v; 21.60.Cs; 
27.20.+n}

\section{Introduction}
Astrophysical processes most often involve charged particle nuclear
reactions at very low energies. The cross sections of such reactions
decrease exponentially with decreasing energy, making their
laboratory measurements extremely difficult. One of the most important
reactions in nuclear astrophysics is the radiative proton capture on
$^7$Be. This reaction produces $^8$B in the Sun, whose $\beta^+$
decay is the main source of high-energy solar neutrinos. That $^7$Be
is radioactive makes the measurement of the low-energy
$^7$Be($p,\gamma$)$^8$B cross section even more difficult. There is a
considerable spread among both the experimental (see, e.g.,
\cite{B8exp}) and theoretical (see, e.g., \cite{B8theo}) values of the
low-energy astrophysical $S_{17}$ factor. However, a common feature in
all of these analyses is the assumption that the cross section at low
energies ($E < 200$~keV) is dominated by the $E1$ transition. Hence
the extrapolated astrophysical $S$ factor is flat.

The $^7$Li$(\vec p,\gamma$)$^8$Be reaction was studied for energies from
0 to 80 keV by Chasteler {\em et al.} \cite{Chasteler} with the hope 
that the somewhat easier experimental situation ($^7$Li is stable) 
might allow one to gain some insight into the similar 
$^7$Be($p,\gamma$)$^8$B process. A considerable forward-backward
asymmetry was found in the angular distribution of the emitted
photons, indicating a non-negligible $M1$ ($p$-wave) strength at such
low energies. A transition-matrix element (TME) analysis of the data
was performed, resulting in two possible values of the $M1$ strength
of 20\% and 80\%. It was suggested that the situation
may be similar in the case of the $^7$Be($p,\gamma$)$^8$B reaction. If
that were so, then the lowest energy ($\approx$ 100 keV)
experimental cross section values would contain a 20\% (80\%) $M1$
contribution. Thus, as the $M1$ cross section becomes negligible at
lower energies because of the centrifugal barrier, the $S$ factor
would not be flat but decreasing with decreasing energy, and the zero
energy $S_{17}$ factor would be 20\% (80\%) smaller than currently
believed.

Rolfs and Kavanagh \cite{RK} suggested that the observed low-energy
$M1$ strength of the $^7$Li($p,\gamma$)$^8$Be reaction comes from the
tail of the $1^+$ state at $E_{lab} = 441$~keV. More recently, the
astrophysical $S$ factor of the $^7$Li$(p,\gamma$)$^8$Be reaction was
accurately measured between 100 and 1500 keV by Zahnow {\em et
al.}  \cite{Zahnow}. The data were interpreted again by assuming that
the small $M1$ contribution comes from the tail of the $1^+$
resonance. Weller and Chasteler \cite{Weller} pointed out that some of
the formulae used in the previous analyses \cite{RK,Zahnow} were
incorrect, and that the resonance tail could only result in a 2\%
$p$-wave contribution, much smaller compared to the earlier
estimates. Barker \cite{Barker} achieved acceptable fits to the
measured angular distributions and analyzing powers by attributing the
$p$-wave $M1$ contribution to the tails of the low-energy
resonances. He obtained an $M1$ strength of $5-6$\%, again much less
than that reported by Chasteler {\em et al.}  However, to achieve
these fits, some of the spectroscopic amplitudes needed the opposite
signs to those obtained from the shell model.

Barker also noted \cite{Barker} that the analogue of the
$^7$Be($p,\gamma$)$^8$B reaction in $^8$Be would be the one which
leads to the $2^+;1$ isobaric analogue state of the ground state of
$^8$B. Such a measurement was performed by Godwin {\em et al.}
\cite{Godwin} and no indication of any low-energy $M1$ strength was
found. Therefore, it may be assumed that the standard extrapolation of the
$^7$Be($p,\gamma$)$^8$B $S$ factor is correct and reliable. However,
the origin of the discrepancies among the various analyses of the
$^7$Li($p,\gamma$)$^8$Be reaction is still not understood. 

Herein, we study the role of the $M1$ transition in the
$^7$Li$(p,\gamma$)$^8$Be reaction by using a microscopic eight-body
model.

\section{Model}

Our model is a microscopic multi-cluster
($^4$He+$^3$H+$p$; $^4$He+$^3$He+$n$; $\alpha$+$\alpha$) Resonating 
Group Method (RGM) approach to the eight-nucleon system. The
trial function of the eight-body system is
\begin{eqnarray}
\Psi&=&\sum_{S,l_1,l_2,L}
{\cal A}\left \{\left [ \left [\Phi^p(\Phi ^\alpha\Phi^t) 
\right ]_S
\chi_{[l_1l_2]L}(\mbox{\boldmath 
$\rho $}_1,\mbox{\boldmath $\rho $}_2)
\right ]_{JM} \right \} \cr
&+&\sum_{S,l_1,l_2,L}{\cal A}\left 
\{\left [ \left [\Phi^n(\Phi ^\alpha\Phi^h) \right ]_S
\chi_{[l_1l_2]L}(\mbox{\boldmath 
$\rho $}_1,\mbox{\boldmath $\rho $}_2)
\right ]_{JM} \right \} \cr
&+&{\cal A}\left \{\left [\Phi ^\alpha\Phi^\alpha 
\chi_L(\mbox{\boldmath 
$\rho $})\right ]_{JM} \right \},
\label{wf}
\end{eqnarray}
where ${\cal A}$ is the intercluster antisymmetrizer, the $\Phi$
cluster internal states are translationally invariant harmonic
oscillator shell model states ($\alpha =$~$^4$He, $t =$~$^3$H, and
$h=$~$^3$He), the \mbox{\boldmath $\rho $} vectors are the different
intercluster Jacobi coordinates, $l_1$ and $l_2$ are the angular
momenta of the two relative motions, $L$ and $S$ are the total orbital
angular momentum and spin, respectively, and [...] denotes angular
momentum coupling. The sum over $S,l_1,l_2$, and $L$ includes all
physically relevant angular momentum configurations. In order to
project out pure $^7$Li and $^7$Be states, the angular momenta are
recoupled in the 7+1 channels following the scheme
\begin{equation}
\bigl [(l_1,l_2)L,(s_1,s_2)S\bigr ]J 
\rightarrow \Bigl [\bigl [(l_1,s_1)I_7,
s_2\bigr ]I,l_2 \Bigr ]J,
\end{equation}
where $I_7$ is the total spin of the seven nucleon system,
and $I$ is the channel spin.

The $^7$Li$(p,\gamma_0$)$^8$Be radiative capture
cross section is calculated perturbatively \cite{pert}, and is 
given by
\begin{eqnarray}
\sigma &=& \sum_{\Omega,\lambda} {{1}\over{(2I_7+1)(2s_2+1)}}
{{8\pi(\lambda +1)}\over{\hbar\lambda(2\lambda +1)!!}} 
\left ({{E_\gamma}\over{\hbar c}}\right )^{2\lambda +1} \cr
&\times& \sum_{l_\omega,I_\omega}(2l_\omega+1)^{-1} 
\vert \langle \Psi^{J_f} \vert \vert {\cal
M}_\lambda^\Omega \vert \vert
\Psi^{J_i}_{l_\omega,I_\omega} \rangle \vert ^2,
\label{sigma}
\end{eqnarray}
where $I_7$ and $s_2$ are the spins of the colliding 
clusters, $\lambda$ denotes the rank of the electromagnetic
operator ${\cal M}_\lambda^\Omega$ ($\Omega=E$ or $M$),
$\omega$ represents the entrance channel, and $J_f$ and
$J_i$ is the total spin of the final and initial state,
respectively. The initial wave function 
$\Psi^{J_i}_{l_\omega,I_\omega}$ is a partial wave of a 
unit-flux scattering wave function. We follow the convention of 
Edmonds \cite{Edmonds} for the reduced matrix elements.

The final $^8$Be ground state is unbound in the $\alpha
+\alpha$ channel which, in principle, could cause
convergence problems. However, this is not the case here
since the $^7$Li+$p$ and $^7$Be+$n$ channel wave
functions are orthogonal to the $\alpha +\alpha$ one in the
asymptotic spatial region. This feature allows us to treat the
$^8$Be ground state as a pseudo-bound state in the $\alpha
+\alpha$ configuration. The orthogonality also indicates
that the $^7$Li+$p$ $\rightarrow$ $\alpha +\alpha$
contribution to the cross section is small, and that the
inclusion of the $7+1$ configurations in the $^8$Be ground
state is important.

Putting (\ref{wf}) into the eight-nucleon Schr\"odinger equation which
contains the nucleon-nucleon ($NN$) strong and Coulomb interactions, we
arrive at an equation for the intercluster relative motion functions
$\chi$. For the $0^+$ bound state these functions are expanded in
terms of products of tempered Gaussian functions $\exp (-\gamma_i
\rho^2)$ \cite{Kamimura} with different ranges $\gamma_i$ for each
type of relative coordinate. The expansion coefficients are determined
from a variational principle. The scattering states are calculated
from a Kohn-Hulth\'en variational method \cite{Kamimura} for the
$S$-matrix, which uses square integrable basis functions matched with
the correct scattering asymptotics. Then, using the resulting
eight-nucleon wave functions, the cross section (\ref{sigma}) is
evaluated.

\section{Results}

The model used herein is essentially the same as that used in
Refs.~\cite{B8,Be7p}, which was used to describe the $^7$Li, $^7$Be,
$^8$Li, and $^8$B nuclei, and the $^7$Be($p,\gamma$)$^8$B reaction. In
those calculations all possible arrangements of the three clusters
were included, e.g., $(\alpha h)p$, $(\alpha p)h$, and $(hp)\alpha$
for $^8$B. However, we consider only the $^7$Li+$p$ and $^7$Be+$n$
configurations due to computational constraints. The results of a full
model and a model which contains only $7+1$ configurations are quite similar
\cite{Be7p}, provided the experimental subsystem properties
(separation energies, channel thresholds, sizes, etc.) are correctly
reproduced in each.

We use the same parameters for the description of $^7$Li and $^7$Be,
and the same $NN$ interaction (Minnesota force) as in Ref.~\cite{B8}.
Using these parameters, the $^7$Li and $^7$Be properties are well
reproduced \cite{B8}. The calculated relative binding energies
$^7$Li(g.s.)--$^7$Be(g.s.) and $^7$Li(g.s.)--$^8$Be(g.s.) are also
close to the experimental values. Note that in eq.~(\ref{sigma})
the experimental value for $E_\gamma$ is used.

We calculate the $E1$, $E2$, and $M1$ cross sections for
the capture to the ground state of $^8$Be. An unambiguous
treatment of the transition to the broad $2^+$ excited
state of $^8$Be would require a more sophisticated model.

As the final state is $J^\pi=0^+$, the $E1$, $E2$, and $M1$
transitions occur from $1^-$, $2^+$, and $1^+$ initial scattering
states, respectively. The $1^-$ and $2^+$ partial waves are
non-resonant above the $^7$Li+$p$ threshold, while there are two $1^+$
resonances at $E_{c.m.}=386$~keV ($\Gamma =11$~keV, $T=1$) and
$E_{c.m.}=897$~keV ($\Gamma = 138$~keV, $T=0$). In order to explore
the role of the $M1$ capture at low energies, the properties of the
386~keV state must be reproduced. We decreased the exchange mixture 
parameter of the $NN$ interaction by 0.4\% and achieved 386 keV 
position and 13 keV width. The
energy and width of the second $1^+$ ($T=0$) state are 1050~keV and
342~keV, respectively, in our model. These resonance parameters were
determined from an analytic continuation of the scattering matrix to
complex energies \cite{He5}. As the $T=0$ state is higher in energy
than experimentally observed, and much broader, it does not have any
observable effect on the low-energy cross section. This is not a
serious problem, as only the first state is expected to contribute
significantly to the cross section at $\approx 80$~keV
\cite{RK,Weller}. Note that the 0.4\% change in the $NN$ interaction
has a negligible effect on the subsystem properties.

We generate the $0^+$ bound state and $1^-$, $2^+$, and
$1^+$ scattering states using the same force in all cases.
In order to account for specific distortions of $^7$Li 
and $^7$Be, we describe these nuclei by using six basis states.
The resulting large number (50--60) of coupled channels
required special numerical attention to keep high precision
in the scattering calculations. 

The result of the calculation of the $^7$Li($p,\gamma$)$^8$Be cross
section is presented in Fig.~\ref{S_log} in terms of the astrophysical
$S$ factor
\begin{equation}
S(E)=\sigma (E)E\exp \Big [2\pi\eta (E) \Big ],
\end{equation}
where $\eta$ is the Sommerfeld parameter. As our $E2$ cross section is
2--3 orders of magnitude smaller than the $E1$, it is not
displayed. There is no contribution to the cross section coming from
the second $1^+$ state, as expected. The nonresonant part of the cross
section and the first $1^+$ state is well reproduced, although some
$M1$ strength is missing in the resonance region.

Note that, as discussed above, inclusion of the $^7$Li+$p$ and
$^7$Be+$n$ configurations in the ground state of $^8$Be is
necessary. The contribution to the cross section from the $\alpha +
\alpha$ final state configuration is very small. This $0^+$ state is
thought to be one of the pure two-cluster ($\alpha+\alpha$) systems
\cite{WT}.  However, by adding the high-lying $7+1$ channels to the
$\alpha+\alpha$ configuration, one gains 1--2~MeV of energy in the
$0^+$ state, indicating that this state may not be a pure
$\alpha+\alpha$ configuration.

In our model the 386 keV ($T=1$) resonance is dominated by the $I=1$ 
channel spin. This is in contradiction to experiment
\cite{Ma60}, which gives 3--5 for the ratio of the $I=2$ to $I=1$
channel spin contributions to the cross section. The same $I=1$
dominance was also observed in the calculations of the $1^+$ isobaric
analogue state in $^8$Li and $^8$B. In order to understand the origin
of this $I=1$ dominance, we calculated the percentages of the
orthogonal $(LS)$ components in $^8$Li (which is the only two-body
bound state in the triplet). The $(LS)=(1,0)$, $(0,1)$, $(1,1)$, and
$(2,1)$ components have 83.4\%, 0.03\%, 13.3\%, and 3.2\% weights,
respectively. However, the contribution of the largest component, 
$(1,0)$, to the $I=2$ channel spin state is zero, and hence the $I=1$
configuration dominates. Very similar $(L,S)$ components were
found in a different three-body model \cite{Shulgina} using completely
different cluster-cluster interactions. That model would also 
predict a dominant $I=1$ component in the $A=8$ $1^+;1$ triplet. In 
the second $1^+$ ($T=0$) state, our model predicts a dominant $I=2$ 
component. It remains a question as to whether the mixing of these 
states could influence the experimental determination of the channel 
spin ratio.

As a comparison to the results obtained from our RGM model, the
$I=1,2$ spectroscopic amplitudes were calculated using wave functions
obtained from a large space shell model calculation. The interaction
used was that of Zheng {\em et al.} \cite{Zh95}, based on a
multi-valued $G$-matrix calculation and defined within the complete
$(0+2+4)\hbar\omega$ shell model space. The calculations were
performed using the shell model code OXBASH \cite{Ox86}. The
spectroscopic amplitudes $S_{I=1,2}$ are defined in terms of the shell
model spectroscopic amplitudes by
\begin{equation}
S_I = \sum_j \alpha_{I,j} \theta_j,
\end{equation}
where $\theta_j$ are the shell model spectroscopic amplitudes ($j=3/2$
or 1/2),
\begin{equation}
\theta_j = \frac{1}{\sqrt{2J_f+1}} \left\langle J_f (\mbox{$^8$Be})
\left\| a^{\dagger}_j \right\| I_7 (\mbox{$^7$Li}) \right\rangle,
\end{equation}
and $\alpha_{I,j}$ are transformation coefficients for the recoupling
of the angular momenta in terms of the channel spin, $I$,
\begin{equation}
\left[ I_7,j \right]J_f \rightarrow \left[ \left( I_7,
\frac{1}{2} \right)I,l_2 \right]J_f.
\end{equation}
Table~\ref{SI} lists the spectroscopic amplitudes, $S_I$, as obtained
from our shell model calculation, along with those obtained from the
calculation using the Cohen and Kurath (CK) $(6-16)$2BME $0p$-shell
interaction \cite{Co65}. Note that the contribution from higher shells
to the amplitudes calculated in the $(0+2+4)\hbar\omega$ space is
negligible. The CK amplitudes favor the $I=2$ channel spin for the
$1^+;1$ state, while the Zheng ones support the results of our
analysis. This discrepancy is the result of a slight redistribution of
particle strength between the $0p_{3/2}$ and $0p_{1/2}$ orbits, as is
illustrated in Table~\ref{Stheta}. The spectroscopic amplitude,
$\theta_j$, for the capture to the $0p_{3/2}$ orbit is larger in the
Zheng model, while there is a corresponding decrease in the magnitude
for the capture to the $0p_{1/2}$ orbit. It is also interesting to
note the signs of the channel-spin amplitudes for the $1^+;0$
state. In Barker's two-level $R$-matrix calculation \cite{Barker}, the
signs of the reduced widths used were obtained from the shell
model. However, Barker changed the sign of the $I=1$ amplitude
relative to that of the $I=2$ one, in order to fit the data. Our shell
model results does not support this sign change.

We note, that if our channel spin ratio is in fact wrong,
then probably the singlet-odd component of the $NN$
force is responsible for this failure. In our model we calculate the
peak to 80 keV $S$ factor ratio for the $I=1$ and $I=2$ channel spin
contributions. Using these numbers we can estimate the $M1$ strength
at 80 keV for any given channel spin ratio.

The necessity of the microscopic description of the
$^7$Li($p,\gamma$)$^8$Be (in contrast to that for
$^7$Be($p,\gamma$)$^8$B \cite{Ba96}, for example) is demonstrated by
noting that in a potential model the $I=2$ contribution would come
almost exclusively from the internal magnetic moments of $^7$Li and
$^7$Be. The orbital part of the $M1$ operator describing the relative
motion (which strongly dominates the spin part) is a rank-zero tensor
in spin space. Thus, for any non-vanishing cross section from the
$^7$Li--$p$ relative motion, the initial and final channel spins must
be equal. Hence, as the channel spin in the $0^+$ is $I=1$, the $I=2$
orbital cross section is zero. In our microscopic model, the $I=2$
cross section comes mainly from nucleon exchange effects.

The $S$ factor is displayed on a linear scale in Fig.~\ref{S_lin}. The
nonresonant $E1$ component is in good agreement with the data. We
emphasize that in our model all parameters are fixed. The requirement
that all relevant properties of the subsystems be reproduced does not
leave any room for adjustment of parameters. 

The contribution of the $M1$ component to the total $S$ factor at
80~keV is 3.5\%. If one adopts a value of 3.2 for the $I=2/I=1$
channel spin ratio, together with the experimental width and peak
value of the 386~keV resonance, this $M1$ contribution is 6.3\%,
consistent with the estimates by Barker \cite{Barker}. The value of
20\% determined from the TME analysis \cite{Chasteler} may be due to
the assumption of a negligible $0p_{3/2}$ spectroscopic factor. That
assumption is not supported by the results of the shell model
calculations presented in Table~\ref{Stheta}.

\section{Conclusions}

In summary, we have studied the low-energy $^7$Li($p,\gamma_0$)$^8$Be
reaction in a microscopic cluster model. All relevant subsystem 
properties are well reproduced by our model. The parameters of the 
$1^+; T=1$ resonance in $^8$Be are in good agreement with the 
experimental values, while the $1^+; T=0$ state is much higher and much
broader in energy compared to experiment.

We have calculated the $E1$, $E2$, and $M1$ cross sections
perturbatively. The nonresonant $E1$ component agrees well with the
experimental data, while the $E2$ contribution is 
negligible. The calculated full astrophysical $S$ factor is in
good agreement with the data of Zahnow {\em et al.} \cite{Zahnow},
except that the $1^+;0$ state does not contribute in our model, and 
some $M1$ strength is missing in the $1^+;1$ resonant region. This is partly
caused by the fact that the width of the $1^+;1$ state is slightly
larger compared to experiment.

Our model predicts that in the $A=8$, $1^+;1$ triplet the dominant
configuration has $I=1$ channel spin. Consequently, the
$^7$Li($p,\gamma_0$)$^8$Be $M1$ cross section is also dominated by the
$I=1$ component at $E_{c.m.}=386$ keV, in contrast with the $I=2$
dominance observed experimentally. Results of large-space shell model
calculations, which give a larger spectroscopic factor for the $I=1$
component in the $1^+; T=1$ state, seem to support our RGM result,
although this depends sensitively on the distribution of strength
between the $0p_{3/2}$ and $0p_{1/2}$ orbits. We note,
however, that the total $M1$ cross section can be estimated for any
channel spin ratio in our model, using the calculated $I=1$ and $I=2$
$M1$ components.

The calculated $M1$ strength at 80 keV is 3.5\% in our model. If we
assume the experimental peak height and width of the 386 keV
resonance in the $S$ factor, and the 3.2 I=2/I=1 channel spin ratio,
the model predicts 6.3\%, in good agreement the result of an $R$
matrix analysis \cite{Barker}.

Our shell model calculations show that the $0p_{3/2}$ $^7$Li+$p$
spectroscopic factor is larger than the $0p_{1/2}$ one in the $1^+;1$
state. Thus the assumption of a vanishing $0p_{3/2}$ spectroscopic
factor in the TME analysis is incorrect. This may have led to the
large $M1$ strength found therein.

\acknowledgements

The work of A.\ C.\ was supported by Wolfgang Bauer's 
Presidential Faculty Fellowship (PHY92-53505). This work
was also supported by NSF Grant No.\ PHY94-03666 (MSU) and  
by the US Department of Energy (Los Alamos). We thank 
Sam Austin, Fred Barker, Alex Brown, and Gregers Hansen 
for useful discussions.

%
%
\begin{figure}
\caption[]{$^7$Li($p,\gamma_0$)$^8$Be astrophysical $S$ factor as a
function of the $^7$Li+$p$ center-of-mass energy. The result of the
present calculation is given by the solid line, while the $E1$ and
$M1$ contributions are given by the dashed and dot-dashed lines,
respectively. The data are those of Zahnow {\em et al.}
\cite{Zahnow}.}
\label{S_log}
\end{figure}
\begin{figure}
\caption{As for Fig.~\ref{S_log}, but displayed on a linear scale.}
\label{S_lin}
\end{figure}

%
%
\begin{table}
\caption[]{Spectroscopic amplitudes, $S_I$, labeled by the channel 
spin $I$ for the $^7$Li($p,\gamma$)$^8$Be reaction to the $1^+$ 
states in $^8$Be, as obtained from the shell model calculations 
using the interactions of Zheng {\em et al.} \cite{Zh95} and of 
Cohen and Kurath \cite{Co65}.}
\label{SI}
\begin{tabular}{cccrr}
      &              &     & \multicolumn{2}{c}{$S_I$} \\
State & $E_{c.m.}$ (keV) & $I$ & Zheng & CK \\
\hline
$1^+;1$ & 386 & 1 & 0.3499 & 0.2644 \\
        &     & 2 & 0.2774 & 0.3910 \\
$1^+;0$ & 897 & 1 & $-$0.2612 & $-$0.2773 \\
        &     & 2 & 0.5132 & 0.5623 \\
\end{tabular}
\end{table}
\begin{table}
\caption[]{Spectroscopic amplitudes, $\theta_j$, for the
$^7$Li($p,\gamma$)$^8$Be reaction to the $1^+$ states in $^8$Be, as
obtained from our shell model calculations.}
\label{Stheta}
\begin{tabular}{cccrr}
      &                  &     & \multicolumn{2}{c}{$\theta_j$} \\
State & $E_{c.m.}$ (keV) & $j$ & Zheng & CK \\
\hline
$1^+;1$ & 386 & $0p_{3/2}$ & 0.4327 & 0.4009 \\
        &     & $0p_{1/2}$ & $-$0.1104 & $-$0.2490 \\
$1^+;0$ & 897 & $0p_{3/2}$ & $-$0.0290 & $-$0.0334 \\
        &     & $0p_{1/2}$ & $-$0.5752 & $-$0.6266 \\
\end{tabular}
\end{table}
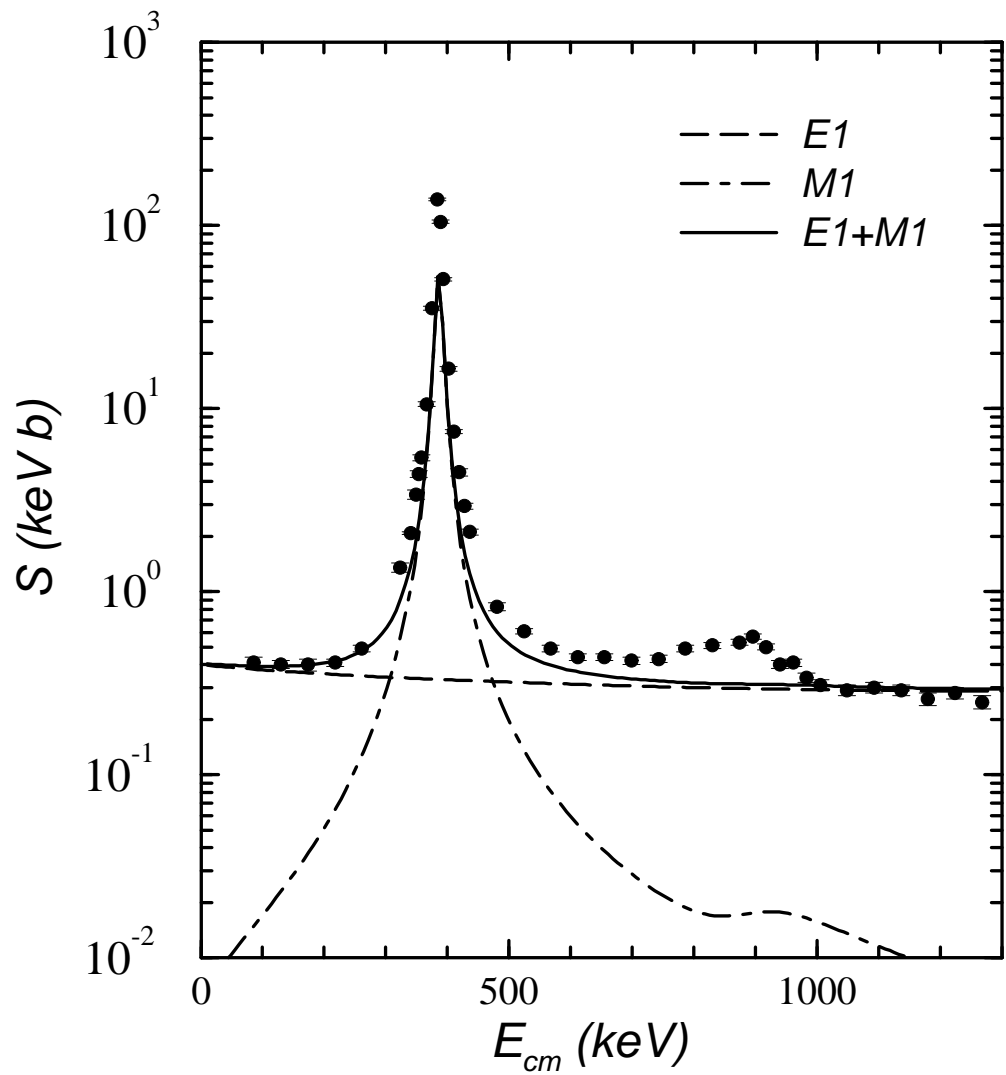
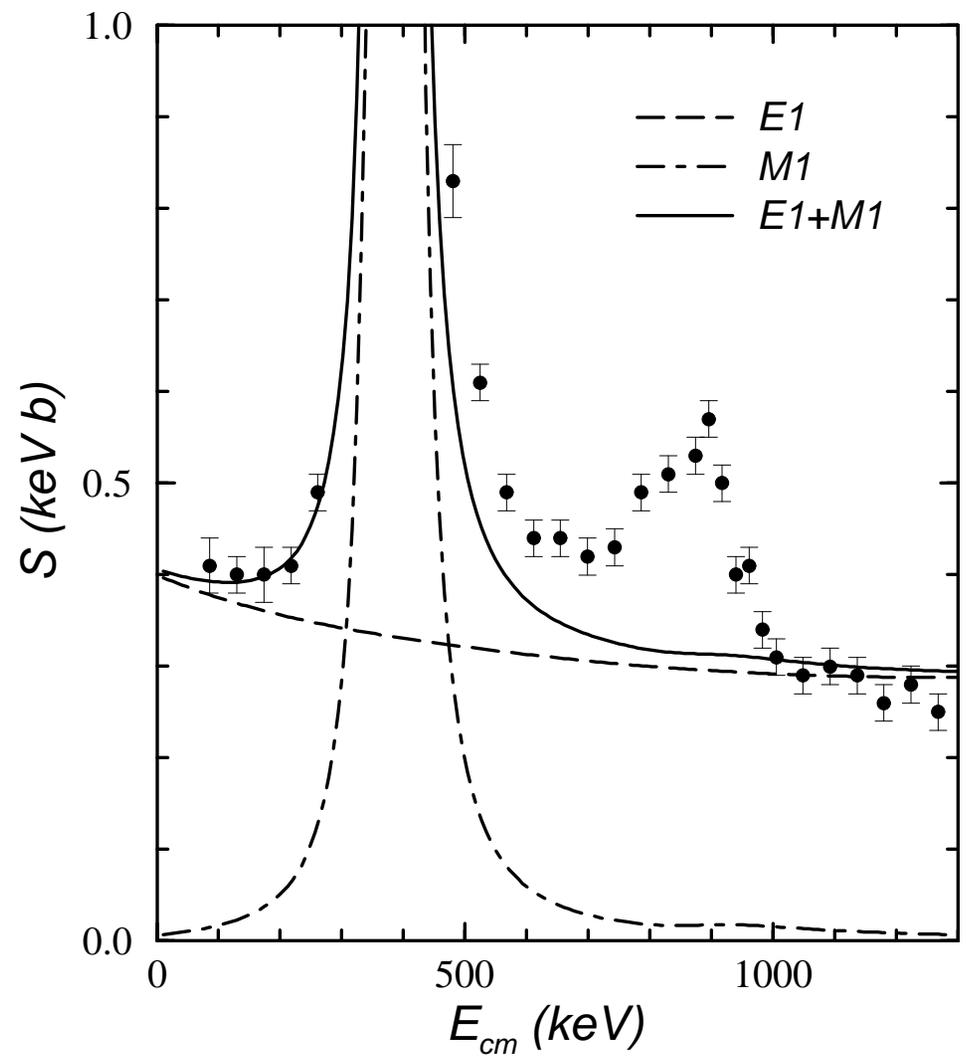
\end{document}